\documentclass[twoside,twocolumn,9pt]{article}
\usepackage{extsizes}
\usepackage[super,sort&compress,comma]{natbib} 
\usepackage[left=1.5cm, right=1.5cm, top=1.785cm, bottom=2.0cm]{geometry}
\usepackage{balance}
\usepackage{mathptmx}
\usepackage{sectsty}
\usepackage{graphicx}
\usepackage{lastpage}
\usepackage[format=plain,justification=justified,singlelinecheck=false,font={stretch=1.125,small,sf},labelfont=bf,labelsep=space]{caption}
\usepackage{float}
\usepackage{fancyhdr}
\usepackage{fnpos}
\usepackage[english]{babel}
\addto{\captionsenglish}{%
  
}
\usepackage{array}
\usepackage{droidsans}
\usepackage{charter}
\usepackage[T1]{fontenc}
\usepackage[usenames,dvipsnames]{xcolor}
\usepackage{setspace}
\usepackage[compact]{titlesec}
\usepackage{hyperref}

\usepackage{chemformula} 
\setchemformula{kroeger-vink}

\usepackage[separate-uncertainty=true,quantity-product = \ ]{siunitx}
\DeclareSIUnit{\kb}{\textit{k}\textsubscript{B}}
\usepackage{chemmacros}
\usepackage{amsmath}

\usepackage{textcomp}

\definecolor{cream}{RGB}{222,217,201}

\begin{document}

\pagestyle{fancy}
\thispagestyle{plain}
\fancypagestyle{plain}{
\renewcommand{\headrulewidth}{0pt}
}

\makeFNbottom
\makeatletter
\renewcommand\LARGE{\@setfontsize\LARGE{15pt}{17}}
\renewcommand\Large{\@setfontsize\Large{12pt}{14}}
\renewcommand\large{\@setfontsize\large{10pt}{12}}
\renewcommand\footnotesize{\@setfontsize\footnotesize{7pt}{10}}
\makeatother

\renewcommand{\thefootnote}{\fnsymbol{footnote}}
\renewcommand\footnoterule{\vspace*{1pt}%
\color{cream}\hrule width 3.5in height 0.4pt \color{black}\vspace*{5pt}} 
\setcounter{secnumdepth}{5}

\makeatletter 
\renewcommand\@biblabel[1]{#1}            
\renewcommand\@makefntext[1]%
{\noindent\makebox[0pt][r]{\@thefnmark\,}#1}
\makeatother 
\renewcommand{\figurename}{\small{Fig.}~}
\sectionfont{\sffamily\Large}
\subsectionfont{\normalsize}
\subsubsectionfont{\bf}
\setstretch{1.125} 
\setlength{\skip\footins}{0.8cm}
\setlength{\footnotesep}{0.25cm}
\setlength{\jot}{10pt}
\titlespacing*{\section}{0pt}{4pt}{4pt}
\titlespacing*{\subsection}{0pt}{15pt}{1pt}

\fancyfoot{}
\fancyfoot[LO,RE]{\vspace{-7.1pt}\includegraphics[height=9pt]{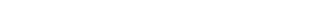}}
\fancyfoot[CO]{\vspace{-7.1pt}\hspace{13.2cm}\includegraphics{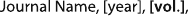}}
\fancyfoot[CE]{\vspace{-7.2pt}\hspace{-14.2cm}\includegraphics{RF}}
\fancyfoot[RO]{\footnotesize{\sffamily{1--\pageref{LastPage} ~\textbar  \hspace{2pt}\thepage}}}
\fancyfoot[LE]{\footnotesize{\sffamily{\thepage~\textbar\hspace{3.45cm} 1--\pageref{LastPage}}}}
\fancyhead{}
\renewcommand{\headrulewidth}{0pt} 
\renewcommand{\footrulewidth}{0pt}
\setlength{\arrayrulewidth}{1pt}
\setlength{\columnsep}{6.5mm}
\setlength\bibsep{1pt}

\makeatletter 
\newlength{\figrulesep} 
\setlength{\figrulesep}{0.5\textfloatsep} 

\newcommand{\topfigrule}{\vspace*{-1pt}%
\noindent{\color{cream}\rule[-\figrulesep]{\columnwidth}{1.5pt}} }

\newcommand{\botfigrule}{\vspace*{-2pt}%
\noindent{\color{cream}\rule[\figrulesep]{\columnwidth}{1.5pt}} }

\newcommand{\dblfigrule}{\vspace*{-1pt}%
\noindent{\color{cream}\rule[-\figrulesep]{\textwidth}{1.5pt}} }

\makeatother

\twocolumn[
  \begin{@twocolumnfalse}
{\includegraphics[height=30pt]{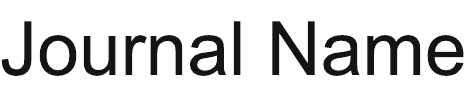}\hfill\raisebox{0pt}[0pt][0pt]{\includegraphics[height=55pt]{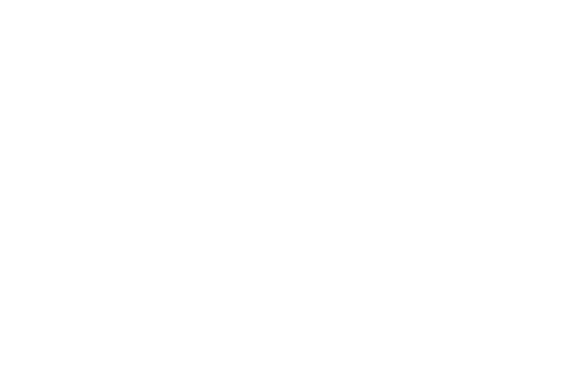}}\\[1ex]
\includegraphics[width=18.5cm]{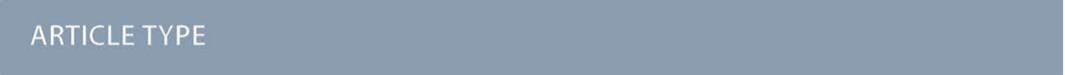}}\par
\vspace{1em}
\sffamily
\begin{tabular}{m{4.5cm} p{13.5cm} }

\includegraphics{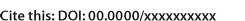} & \noindent\LARGE{\textbf{Faster grain-boundary diffusion with a higher activation enthalpy than bulk diffusion in ionic space-charge layers}} \\
\vspace{0.3cm} & \vspace{0.3cm} \\

 & \noindent\large{Timon F. Kielgas,\textit{$^{a}$} Roger A. De Souza$^{\ast}$\textit{$^{a}$}} \\

\includegraphics{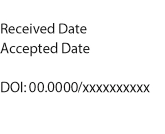} & \noindent\normalsize{Faster diffusion of cations along grain boundaries is reported in the literature for a variety of acceptor-doped $AB\mathrm{O}_{3}$ perovskite-type oxides. The ratio $r$ of the activation enthalpy of grain-boundary diffusion ($\Delta H^\mathrm{gb}$) to the activation enthalpy of bulk diffusion ($\Delta H^\mathrm{b}$) is seen experimentally to lie in the range $0.7 < r = \Delta H^\mathrm{gb} / \Delta H^\mathrm{b} < 1.3$, albeit with substantial errors. In a previous publication [Parras and De Souza, Acta Mater. 195 (2020) 383] it was shown through a set of continuum simulations that cation-vacancy accumulation within negative space-charge layers at grain boundaries in acceptor-doped perovskites will give rise to faster grain-boundary diffusion of cations, with the associated values of $r$ approaching but not exceeding unity. In the present study, we demonstrate by means of continuum simulations that $r > 1$ is possible for faster cation diffusion along grain boundaries in an acceptor-doped perovskite. The specific case we consider is 
cation diffusion occurring by two related mechanisms, by slower (charged) isolated cation vacancies and by faster (neutral) defect associates of cation and anion vacancies. Within the negative space-charge layers, the isolated cation vacancies are strongly accumulated, whereas the neutral associates are unaffected. We calculate diffusion profiles for a two-dimensional bicrystal geometry by solving, first, Poisson's equation, and subsequently, the diffusion equation. We find that, if a small concentration of faster defect associates is responsible for bulk diffusion, and a hugely enhanced concentration of slower isolated vacancies yields faster diffusion along space-charge layers, $r>1$ is obtained. The conditions under which $r > 1$ may be observed are described, and issues with experimental confirmation are discussed.
} \\ 


\end{tabular}

 \end{@twocolumnfalse} \vspace{0.6cm}

  ]

\renewcommand*\rmdefault{bch}\normalfont\upshape
\rmfamily
\section*{}
\vspace{-1cm}


\footnotetext{\textit{$^{a}$~Institute of Physical Chemistry, RWTH Aachen University, Landoltweg 2, 52074 Aachen, Germany. Fax: +49 241 80 92128; Tel: +49 241 80 94739; E-mail: desouza@pc.rwth-aachen.de,  kielgas@pc.rwth-aachen.de}}


\section{Introduction}

In ionic solids, grain boundaries, as two-dimensional defects, are not required in electrochemical equilibrium to be locally neutral. Grain boundaries are expected to be electrostatically charged and thus enveloped in compensating space-charge layers to fulfil global electronetrality. A neutral grain boundary is expected to be the exception. Within the space-charge layers, the concentrations of charged point defects may differ by orders of magnitude from their bulk values, with consequences for those processes that depend directly on defects, such as the transport of ions.

Prime examples of ionic solids with charged grain boundaries are acceptor-doped and acceptor-substituted titanate perovskites. The transport of oxide ions across pristine grain boundaries (i.e.~those free of second phases) is known to be hindered.\cite{Schaffrin.1976,Watanabe.2013,Leonhardt.1999,Waser.1989,Denk.1997,DeSouza.2003,DeSouza.2014} This is attributed to the grain boundaries being positively charged, which leads to the positively charged oxygen vacancies (the defects responsible for oxygen diffusion in these materials) being depleted in the neighboring space-charge layer.\cite{DeSouza.2009,Usler.2024,Denk.1997,McIntyre.2000,Waser.1995,Waser.2000}

Much less attention, however, has been devoted to how space-charge layers affect mass transport along grain boundaries in these materials. The species that are expected to diffuse at faster rates along space-charge layers than in the bulk phase are cations, since the defects responsible for cation diffusion in perovskites are negatively charged cation vacancies, and these defects will be strongly accumulated in the space-charge layers.\cite{Usler.2021,Usler.2024,GarciaBelmonte.2025}

\begin{figure}[t]
  \centering
  \includegraphics[width=\columnwidth]{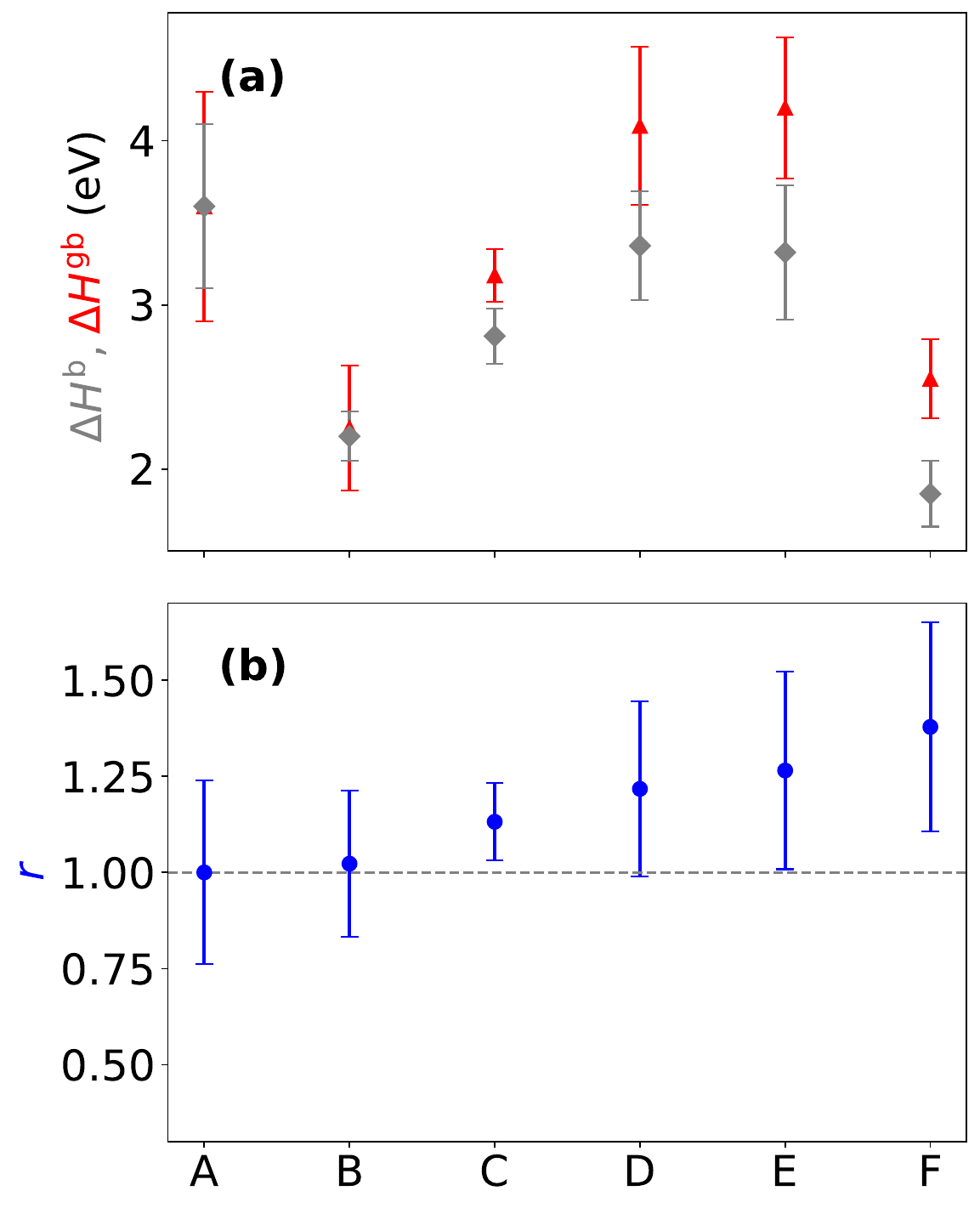}
  \caption{Cation diffusion in selected $AB\mathrm{O}_3$ perovskite oxides: (a) Activation enthalpies for bulk diffusion $\Delta H^\mathrm{b}$ and for the grain-boundary diffusion product $\Delta H^\mathrm{gb}$. (b) The ratio of the activation enthalpies, $r=\Delta H^\mathrm{gb}/\Delta H^\mathrm{b}$. A: Ce in \ch{BaZrO3} \cite{Hasle.2021}  B: \ch{^{52}Cr} in \ch{La_{0.8}Sr_{0.2}Ga_{0.8}Mg_{0.2}O_{2.8}} \cite{O.Schulz.2003}, C: \ch{^{56}Fe} in \ch{LaGaO3} \cite{O.Schulz.2003}, D: Fe-Co interdiffusion in \ch{La_{0.6}Sr_{0.4}CoO_{3-$\delta$}} \cite{Kubicek.2014} E: \ch{^{86}Sr} in \ch{La_{0.6}Sr_{0.4}CoO_{3-$\delta$}} \cite{Kubicek.2014} F: \ch{^{56}Fe} in \ch{La_{0.8}Sr_{0.2}Ga_{0.8}Mg_{0.2}O_{2.8}}\cite{O.Schulz.2003}}
  \label{fig:Lit_activation_r_overview}
\end{figure}

Experimentally, faster diffusion of cations along grain boundaries in acceptor-doped perovskite oxides has indeed been observed,\cite{Parras.2020,Wrnhus.2007,Sazinas.2017b,Sazinas.2017,Palcut.2008,Yamazaki.2000,Itoh.2004,Smith.2006,Harvey.2012,Kubicek.2014} but altered point-defect concentrations in space-charge layers are rarely invoked as an explanation. Generally, the observed behaviour is attributed, as in metals, to the atomic arrangement within the grain-boundary core being more open than in the bulk. As in metals, the activation enthalpy of grain-boundary diffusion is expected, therefore, to be roughly half as high as that of bulk diffusion, that is, $r = \Delta H^\mathrm{gb}/\Delta H^\mathrm{b} \approx 0.5$, and thus diffusion, being exponentially dependent on the activation enthalpy, is much faster along grain boundaries than in the bulk, \hbox{$D^\mathrm{gb} \gg D^\mathrm{b}$.} There is a problem, however, in applying this standard picture of faster grain-boundary diffusion to the diffusion of cations in acceptor-doped perovskite oxides. If one examines values of $r$ obtained experimentally (see Fig.~\ref{fig:Lit_activation_r_overview}), one generally finds values well above the $r\approx 0.5$ expected for metals, and in fact, values closer to, and even exceeding, unity. Such values are physically inconsistent with the standard picture, and thus suggest that there is a phenomenon in oxides, but not in metals, that generates $D^\mathrm{gb} \gg D^\mathrm{b}$. The one key difference between metals and oxides, as pointed out by Kingery,\cite{KINGERY.1974} is that the latter are composed of ions, and as a consequence, space-charge zones may develop at extended defects.

Starting from a thermodynamic description of space-charge layer formation,\cite{McIntyre.2000,Tschoepe.2004,DeSouza.2009} Parras and De Souza used continuum simulations to examine values of $r$ for the case of faster diffusion along space-charge layers at a grain boundary (but not within the core). Their results indicated $r \leq 1$, and they presented quantitative arguments that ruled out higher values of $r$ within their model (of a dilute solution of point defects and diffusion occurring by a single migration mechanism).\cite{Parras.2020} The question thus arises whether $r > 1$ is due to experimental error or whether, within a different model, it is physically reasonable.

In this study, we used continuum simulations to examine if $r > 1$ is physically possible for faster cation diffusion along grain boundaries in acceptor-doped perovskite oxides. To this end, we relaxed to a limited degree the constraint (used by Parras and De Souza\cite{Parras.2020}) that the migration mechanism is the same in the bulk and in the space-charge layers. The possibility of two completely different mechanisms (e.g.~vacancy migration in the bulk and interstitial migration in the space-charge layers) is ignored because it is not applicable to cation diffusion in perovskites (the energies of Frenkel disorder being prohibitively high\cite{Liu.2014,Akhtar.1995,Thomas.2007,Stokes.2010,DeSouza.2003b,DeSouza.1999}), and further, because it is a somewhat trivial case. Rather, being aware of two particular, related strontium-vacancy migration mechanisms in \ch{SrTiO3},\cite{Walsh.2011,Heelweg.2021,Gries.2020} and being aware of the positive space-charge potential at grain boundaries in acceptor-doped perovskites,\cite{DeSouza.2009,Usler.2024,Denk.1997,McIntyre.2000,Waser.1995,Waser.2000} we considered the possibility of faster diffusion of \ch{Sr} along a grain boundary in \ch{SrTiO3}. Specifically, the two related mechanisms are the migration of slower (charged) isolated strontium vacancies (with an activation enthalpy of ca.~\qty{4}{\electronvolt}) and of faster (neutral) defect associates of strontium and oxygen vacancies (with an activation enthalpy of ca.~\qty{3.4}{\electronvolt}).\cite{Walsh.2011,Heelweg.2021,Gries.2020} And specifically, the positive space-charge potential causes the negatively charged, isolated cation vacancies to be strongly accumulated, whilst not affecting the neutral associates. Thus, if the defect associates form in sufficient numbers in the bulk, they may, with their lower migration enthalpy, dominate bulk diffusion, whereas within the space-charge layers, the enhanced concentration of isolated cation vacancies may—despite their higher migration enthalpy—give rise to faster grain-boundary diffusion. 


In order to implement this model, we take a similar approach to Parras and De Souza:\cite{Parras.2020} starting from a thermodynamic description of space-charge formation, using a dilute-solution model of \ch{SrTiO3}'s defect chemistry, and focussing on a bicrystal geometry, with the grain boundary situated normal to the surface. We computed point-defect concentrations, and subsequently, \ch{Sr} cation diffusion profiles in the simulation cell by finite-element-method (FEM) calculations. From these results, we extract the grain-boundary diffusion product $a^\mathrm{gb}D^\mathrm{gb}$ and from data as function of temperature, the corresponding activation enthalpy for the grain-boundary diffusion of \ch{Sr}, $\Delta H^\mathrm{gb}$, which we then compare with the bulk activation enthalpy $\Delta H^\mathrm{b}$. 

\section{Point-defect concentrations in \ch{SrTiO3}} \label{sec:Defect_model_SrTiO3}

\subsection{Bulk-defect Chemistry} \label{sec:Defect_model_SrTiO3_bulk_defects}

The defect chemistry of \ch{SrTiO3} is well established,\cite{Balachandran.1981,Chan.1981,Choi.1986,CHOI.1988,Waser.1991,Denk.1995,Moos.1997} and for the temperatures of interest ($T > \qty{1200}{\kelvin}$), one needs to take the partial Schottky equilibrium for \ch{SrO} into account (as well as the generation of electron and holes, and the reduction of the oxide). In the present case, this standard defect model needs only to be extended by considering the association of strontium vacancies and oxygen vacancies. All defects are assumed to be fully ionised for all (temperature $T$ and oxygen activity $a\mathrm{O}_2$) conditions considered. Defect ionisation equilibria (for acceptor dopants, for strontium vacancies, for oxygen vacancies and for associates) are not included in the model, since their inclusion, for the purposes of this study, would add unnecessary complexity to the defect behaviour in the bulk and in the space-charge layers.

\begin{figure*}[h!]
  \centering
  \includegraphics[width=\textwidth]{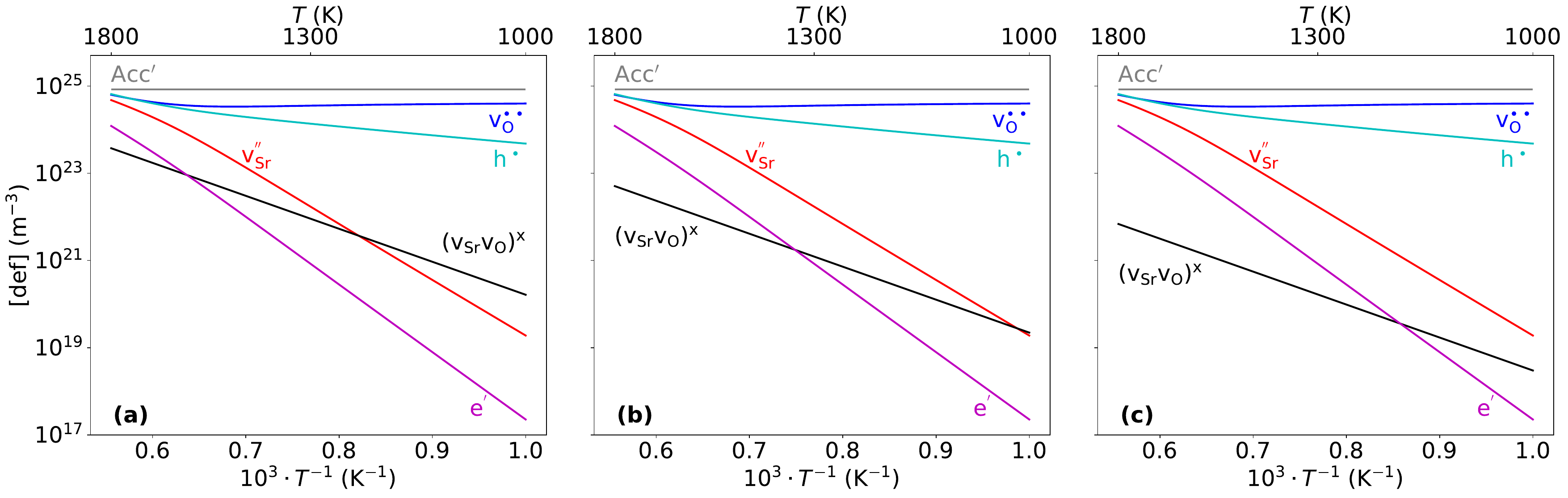}
  \caption{Bulk concentrations of point defects in acceptor-doped \ch{SrTiO3} as a function of inverse temperature. Entropy of associate formation is varied: (a)~\(\Delta S_{\text{a}} = \qty{0}{\kb}\), (b) \(\Delta S_{\text{a}} = \qty{-2}{\kb}\), (c) \(\Delta S_{\text{a}} = \qty{-4}{\kb}\). Parameters are \([\ch{Acc'}] = \qty{8.43e24}{\per\cubic\meter}\), $a\mathrm{O_2} = 0.2$, and \(\Delta H_\mathrm{a} = \qty{-1}{\electronvolt}\).}
  \label{fig:bulk_composition_v4}
\end{figure*}

The defect-chemical equilibrium describing the association of strontium vacancies and oxygen vacancies,\cite{Gries.2020,Heelweg.2021} in Kröger--Vink notation,\cite{F.A.Kroger.1956,Norby.2010,DeSouza.2023} reads as
\begin{equation}
  \ch{v_{Sr}^{''} + v_{O}^{**} <=> (v_{Sr}v_{O})^x}, \label{eq:associate_formation}
\end{equation}
with equilibrium constant 
\begin{equation}
  K_\mathrm{a} = \dfrac{[\ch{(v_{Sr}v_{O}^{x})}]}{[\ch{v_{Sr}^{''}}] \cdot [\ch{v_{O}^{**}}]} = K_\mathrm{a}^{0} \cdot \exp\left( \dfrac{\Delta S_\mathrm{a}}{k_\mathrm{B}} \right) \cdot \exp\left( -\dfrac{\Delta H_\mathrm{a}}{k_\mathrm{B}T} \right),\label{eq:associate_formation_equilibrium}
\end{equation}
where $[\mathrm{def}]$ denotes the concentration of species $\mathrm{def}$ in \unit{\per\cubic\meter}; $K_\mathrm{a}^{0}$ is the pre-exponential term; $\Delta H_\mathrm{a}$ and $\Delta S_\mathrm{a}$ are the entropy and enthalpy of associate formation. 

We consider acceptor-doped \ch{SrTiO3}, with an acceptor dopant of constant valence. The electroneutrality condition for this model is thus
\begin{equation}
  \ch{2 [v_{O}^{**}] + [h^{*}] = [e^{'}] + 2 [v_{Sr}^{''}] + [Acc'] }. \label{eq:electroneutrality_schottky}
\end{equation}
The other defect-chemical equilibria are, as mentioned above, the \ch{SrO} partial-Schottky equilibrium,
\begin{equation}
    \ch{ O_{O}^{x} + Sr_{Sr}^{x} <=> v_{Sr}^{''} + v_{O}^{**} + SrO }, \label{eq:reac_schottky_defect} 
\end{equation}  
with equilibrium constant
\begin{equation}
    K_\mathrm{Sch} = [\ch{v_{Sr}^{''}}] \cdot [\ch{v_{O}^{**}}] = K_\mathrm{Sch}^{0} \cdot \exp\left( -\frac{\Delta H_\mathrm{Sch}}{k_\mathrm{b}T} \right); \label{eq:K_schottky_defect} 
\end{equation}
the oxygen reduction reaction,
\begin{equation}
    \ch{ O_{O}^{x} <=> v_{O}^{**} + 2 e^{'} + 1/2 O2 }, \label{eq:reac_oxygen_incorporation}
\end{equation}  
with equilibrium constant
\begin{equation}
    K_\mathrm{Red} = [\ch{v_{O}^{**}}] \cdot [\ch{v_{e}^{'}}]^{2} \cdot a_{\ch{O2}}^{0.5} = K_\mathrm{Red}^{0} \cdot \exp\left( { -\frac{\Delta H_\mathrm{Red}}{k_\mathrm{b}T} }\right); \label{eq:K_oxygen:_incorporation}
\end{equation}
and the thermal excitation of electrons across the band gap to form conduction-band electrons and valence-band holes.
\begin{equation}
    \ch{nil <=> e^{'} + h^{*}}, \label{eq:reac_e_h_formation}
\end{equation} 
with equilibrium constant
\begin{equation}
    K_\mathrm{eh} = [\ch{e^{'}}] \cdot [\ch{h^{*}}] = N_\mathrm{C}(T) N_\mathrm{V}(T) \exp\left( -\frac{E_\mathrm{g}^{0} - \beta_\mathrm{g}T}{k_\mathrm{b}T} \right), \label{eq:K_e_h_formation}
\end{equation}

\noindent with the effective density of states in the conduction and valance band \hbox{$N_\mathrm{C}(T)$}, \hbox{$N_\mathrm{V}(T)$}, the band gap \hbox{$E_\mathrm{g}^{0}$}, and the temperature coefficient \hbox{$\beta_\mathrm{g}$}. Values for $K_\mathrm{Sch}$, $K_\mathrm{Red}$, $K_\mathrm{eh}$ were calculated from the parameters derived by Moos and Härdtl,\cite{Moos.1997} and are listed in Table \ref{tbl:defect_model_parameters}. The association reaction has not been examined experimentally, and hence a value of $\Delta H_\mathrm{a} = \qty{-1}{\electronvolt}$ was used; $\Delta S_\mathrm{a}$ was taken to have a range of values, $\Delta S_\mathrm{a} / \unit{\kb} = \{0, -2, -4, -6, -8, -10\}$, in order to shift the temperature at which the association equilibrium is predominantly on the right-hand side or the left-hand side of eqn \eqref{eq:associate_formation}.

\begin{table}[h!]
\small
  \caption{\ Parameters used in this study to model the defect chemistry and the diffusion of Sr in bulk \ch{SrTiO3}.}
  \label{tbl:defect_model_parameters}
  \begin{tabular*}{0.48\textwidth}{@{\extracolsep{\fill}}lllr}
    \hline
    Parameter & Value & Units & Ref. \\
    \hline
    $K_\mathrm{Sch}^{0}$ & \num{3e56} & \unit{\meter\tothe{-6}} & \citenum{Moos.1997} \\
    $\Delta H_\mathrm{Sch} $ &  \num{2.5} & \unit{\electronvolt} & \citenum{Moos.1997} \\
    $K_\mathrm{Red}^{0}$ & \num{5e89} & \unit{\meter\tothe{-9}\bar\tothe{0.5}} & \citenum{Moos.1997}\\
    $\Delta H_\mathrm{Red} $ &  \num{6.1} & \unit{\electronvolt} & \citenum{Moos.1997}\\
    $N_\mathrm{C}(T)$ & \num{4.1e22} $(T/\mathrm{K})^{1.5}$ & \unit{\per\cubic\meter} & \citenum{Moos.1997} \\
    $N_\mathrm{V}(T)$ & \num{3.5e22} $(T/\mathrm{K})^{1.5}$ & \unit{\per\cubic\meter} & \citenum{Moos.1997}\\
    $E_\mathrm{g}^{0}$ & \num{3.17} & \unit{\electronvolt} & \citenum{Moos.1997}\\
    $\beta_\mathrm{g}$ & \num{5.66e-4} & \unit{\electronvolt\per\kelvin} & \citenum{Moos.1997}\\
    $K_\mathrm{a}^0$ & \num{1.98e-29} & \unit{\cubic\meter} & -- \\
    $\Delta S_\mathrm{a}$ & $\{0, -2, -4, -6, -8, -10\}$ & \unit{\kb} & -- \\
    $\Delta H_\mathrm{a}$ & \num{-1} & \unit{\electronvolt} & -- \\
    $d_\mathrm{Sr} $ & \num{3.905e-10} & \unit{\meter} & -- \\
    $\nu_{0, i}$ & \num{10.2e12} & \unit{\hertz} & \citenum{Heelweg.2021} \\
    $\Delta S_\mathrm{mig, i}^{\ddagger}$ & \num{2.0} & \unit{\kb} & \citenum{Heelweg.2021} \\
    $\Delta H_\mathrm{mig, i}^{\ddagger}$ & \num{3.96} & \unit{\electronvolt} & \citenum{Heelweg.2021} \\
    $\nu_{0, a}$ & \num{20.3e12} & \unit{\hertz} & \citenum{Heelweg.2021} \\
    $\Delta S_\mathrm{mig, a}^{\ddagger}$ & \num{0.88} & \unit{\kb} & \citenum{Heelweg.2021} \\
    $\Delta H_\mathrm{mig, a}^{\ddagger}$ & \num{3.41} & \unit{\electronvolt} & \citenum{Heelweg.2021} \\
    \hline
  \end{tabular*}
\end{table}

Eqns \eqref{eq:associate_formation_equilibrium}, \eqref{eq:electroneutrality_schottky}, \eqref{eq:K_schottky_defect}, \eqref{eq:K_oxygen:_incorporation} and \eqref{eq:K_e_h_formation} were solved simultaneously with the Python package \textsc{massaction}\cite{Usler.2024b} to yield defect concentrations as a function of $T$ for $a\mathrm{O}_2 = 0.2$ and [\ch{Acc'}] = \qty{8.43e24}{\per\cubic\meter} (dopant site fraction of 0.05 \%). Results are plotted in Fig.~\ref{fig:bulk_composition_v4} for three selected values of $\Delta S_\mathrm{a}$. While appreciable cation diffusion in the bulk is not expected for $T < \qty{1200}{\kelvin}$,\cite{Gomann.2005,Gries.2020,Meyer.2003} it is useful here to consider data below this temperature. Anticipating later results, we make three remarks on the general behaviour shown in Fig. \ref{fig:bulk_composition_v4}. First, the temperature at which the isolated and associated vacancies are equal shifts to lower temperatures with decreasing $\Delta S_\mathrm{a}$. Second, $[\ch{v_{Sr}^{''}}]$ remains unaffected by the association equilibrium, even when $[\ch{(v_{Sr}v_{O})^{x}}] > [\ch{v_{Sr}^{''}}]$ [Fig.~\eqref{fig:bulk_composition_v4} (a)], since $[\ch{v_{O}^{**}}]$ is so high that it fixes $[\ch{v_{Sr}^{''}}]$ through $K_\mathrm{Sch}$ [eqn \eqref{eq:K_schottky_defect}]. Third, [\ch{v_{O}^{**}}] exhibits in all cases a small but clear minimum, and this occurs when $[\ch{h^{*}}]$ and  $[\ch{v_{Sr}^{''}}]$ become appreciable relative to $[\ch{Acc{'}}]$ and $[\ch{v_{O}^{**}}]$.

Strontium ions in perovskite \ch{SrTiO3} can, as mentioned above, diffuse by means of isolated as well as by means of associated vacancies,\cite{Walsh.2011,Heelweg.2021,Gries.2020} and the respective diffusion coefficients are $D_\mathrm{i}$ and $D_\mathrm{a}$. The total diffusivity of strontium cations in the bulk is assumed to be the sum of the two, weighted by their respective site fraction.

\begin{equation}
  D_\mathrm{Sr} = D_\mathrm{i} \cdot \dfrac{[\ch{v_{Sr}^{''}}]}{[\ch{Sr_{Sr}^{x}}]} + D_\mathrm{a} \cdot \dfrac{[\ch{(v_{Sr}v_{O})^{x}}]}{[\ch{Sr_{Sr}^{x}}]}. \label{eq:diffusivity_sum}
\end{equation}

\noindent Each defect diffusion coefficient $D_\mathrm{def}$ is given by 
\begin{equation}
  D_\mathrm{def} = \dfrac{Z_\mathrm{def}}{6} d^2_\mathrm{Sr} \nu_{0, \mathrm{def}} \cdot \exp\left( \dfrac{\Delta S_\mathrm{mig, def}^{\ddagger}}{k_\mathrm{B}} \right) \cdot \exp\left( -\dfrac{\Delta H_\mathrm{mig, def}^{\ddagger}}{k_\mathrm{B}T} \right). \label{eq:diffusivity_vacancy}
\end{equation}
$Z_\mathrm{def}$ is the number of jump neighbors ($Z_\mathrm{i} = 6$, $Z_\mathrm{a} = 2$); $d_\mathrm{Sr}$, the jump distance; $\nu_\mathrm{0, def}$, a characteristic lattice frequency; and $\Delta S_\mathrm{mig, def}^{\ddagger}$ and $\Delta H_\mathrm{mig, def}^{\ddagger}$, the activation entropy and enthalpy of defect migration (see Table~\ref{tbl:defect_model_parameters}).

\begin{figure}[h!]
  \centering
  \includegraphics[width=1.0\columnwidth]{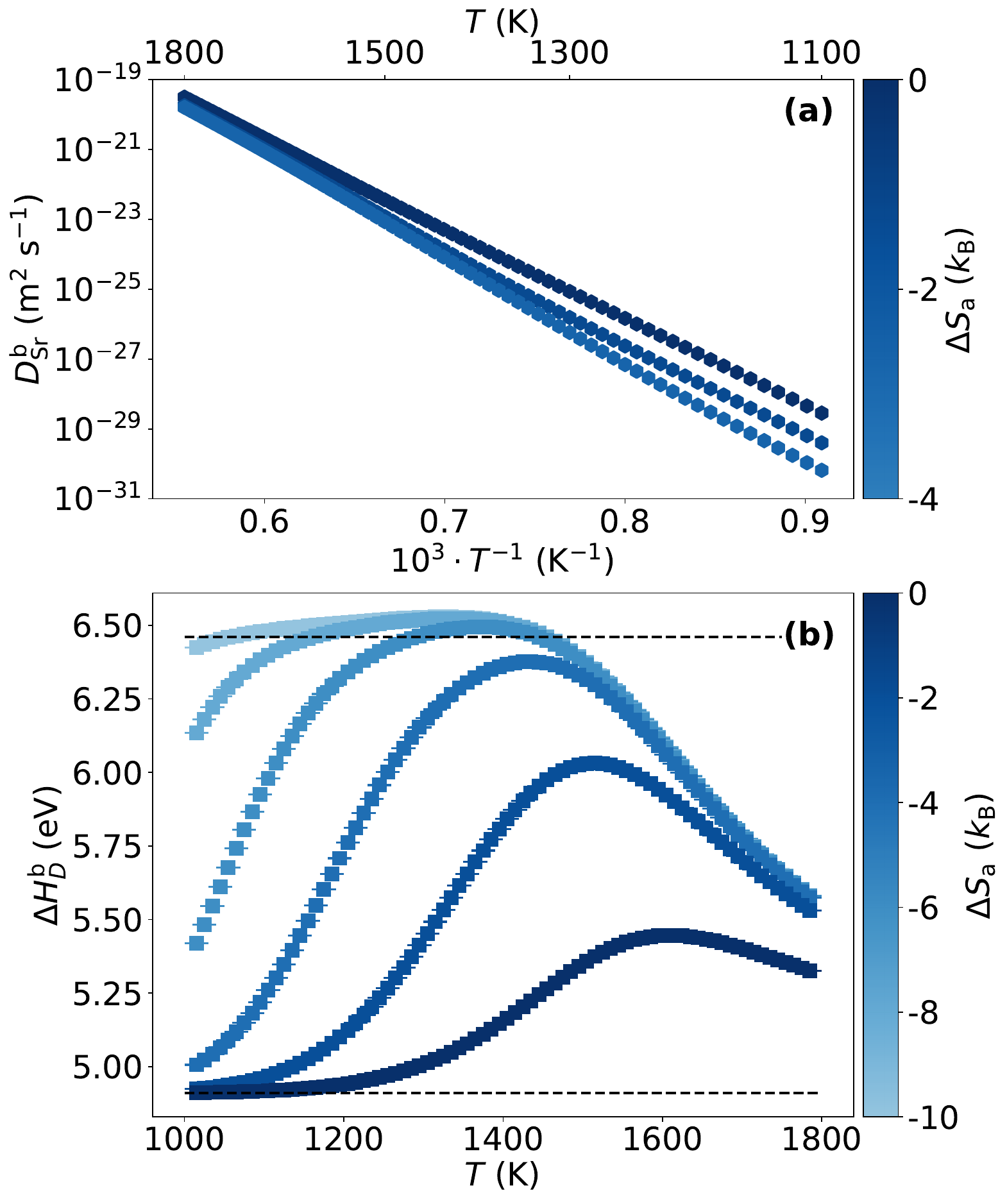}
  \caption{Bulk diffusion coefficients of strontium in \ch{SrTiO3} calculated from defect concentrations and defect diffusivities for diffusion by means of isolated vacancies and defect associates. (a) Bulk diffusion coefficients as a function of inverse temperature. (b) Activation enthalpy of bulk diffusion over a rolling interval of \qty{40}{\kelvin}, plotted as a function of temperature. Different datasets refer to differing values of the entropy of defect association $\Delta S_\mathrm{a}$.}
  \label{fig:full_model_D_and_H}
\end{figure}

Eqns \eqref{eq:diffusivity_sum} and \eqref{eq:diffusivity_vacancy} were then used, together with the calculated defect concentrations, to obtain $D^\mathrm{b}_\mathrm{Sr} (T)$ for different entropies of associate formation. The results are shown in \hbox{Fig.~\ref{fig:full_model_D_and_H} (a).} $D^\mathrm{b}_\mathrm{Sr}$ for different $\Delta S_\mathrm{a}$ converge at higher temperatures, since isolated $\ch{v_{Sr}^{''}}$ dominates over associates (cf.~Fig.~\ref{fig:bulk_composition_v4}). At lower temperatures, $D^\mathrm{b}_\mathrm{Sr}$ is larger for smaller values of $\Delta S_\mathrm{a}$, since this yields higher concentrations of the more mobile $\ch{(v_{Sr}v_{O})^{x}}$ associate.

From the $D^\mathrm{b}_\mathrm{Sr} (T)$ data, the activation enthalpy of strontium diffusion in the bulk is calculated for a rolling average of $\qty{40}{\kelvin}$ [Fig.~\ref{fig:full_model_D_and_H} (b)]. For all values of $\Delta S_\mathrm{a}$ considered, $\Delta H^\mathrm{b}_{D}$ goes through a maximum. This behaviour can, of course, be understood in terms of the defect chemistry and the defect diffusivities. For lower temperatures, the electroneutrality condition can be approximated by \hbox{$\ch{[Acc']} = 2[\ch{v_{O}^{**}}]$.} If the associates provide the dominant contribution to $D_\mathrm{Sr}^\mathrm{b}$, then the activation enthalpy of strontium diffusion can be shown to be \hbox{$\Delta H^\mathrm{b}_{D} = \Delta H^\mathrm{\ddagger}_\mathrm{mig, a} + \Delta H_\mathrm{a} + \Delta H_\mathrm{Sch} = \qty{4.91}{\electronvolt}$.} For intermediate temperatures, with a dominant contribution from isolated strontium vacancies, and with the same approximate electroneutrality condition, one obtains \hbox{$\Delta H^\mathrm{b}_{D} = \Delta H^\mathrm{\ddagger}_\mathrm{mig, i}  + \Delta H_\mathrm{Sch} = \qty{6.46}{\electronvolt}$.} These two values are indicated by dashed lines in Fig.~\ref{fig:full_model_D_and_H}(b). The higher limit is exceeded slightly by the actual data for datasets with higher $\Delta S_\mathrm{a}$ because the contribution of the associates to $D^\mathrm{b}_\mathrm{Sr}$ is not negligible. And finally, at the highest temperatures considered, isolated strontium vacancies continue to provide the dominant contribution, but the electroneutrality approximation effectively becomes \hbox{$2[\ch{v_{Sr}^{''}}] + [\ch{Acc^{'}}] = 2[\ch{v_{O}^{**}}] + [\ch{h^{*}}]$.} In this case no simple expression can be derived for $\Delta H^\mathrm{b}_{D}$. Nevertheless, the data in Fig.~\ref{fig:bulk_composition_v4} indicates that the temperature dependence of [\ch{v_{Sr}^{''}}] becomes weaker with the change in electroneutrality condition, and hence, relative to the data below this temperature, $\Delta H^\mathrm{b}_{D}$ has to decrease.

Since experimental studies of cation diffusion in perovskites are extremely difficult to carry out, and the density of experimental data points as a function of temperature is accordingly rather low, possible deviations from linear Arrhenius behaviour may rarely be directly evident. 

\begin{figure*}[h!]
  \centering
  \includegraphics[width=1.0\textwidth]{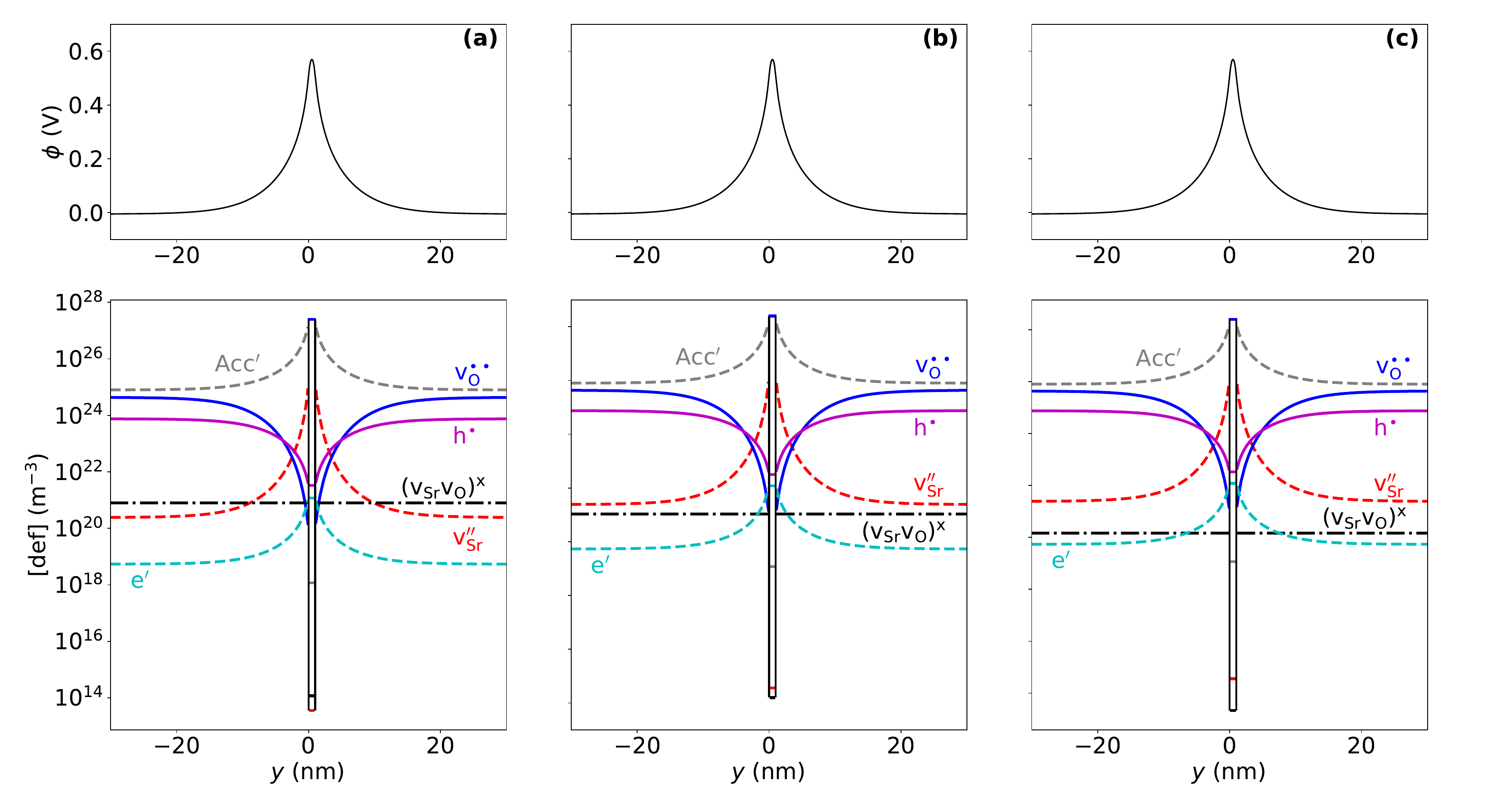}
  \caption{Electric potential $\phi$ and defect concentrations [def] across a grain boundary, for which oxygen-vacancy segregation ($\Delta \mu_\mathrm{v}^\standardstate = \qty{-2}{\electronvolt}$) drives space-charge formation, producing a positively charged core and negative space-charge layers. Positive defects are plotted with solid lines; negative defects, with dashed lines; and the neutral defect, with a dot-dashed line. Entropy of associate formation is varied: (a)~\(\Delta S_{\text{a}} = \qty{0}{\kb}\), (b) \(\Delta S_{\text{a}} = \qty{-2}{\kb}\), (c) \(\Delta S_{\text{a}} = \qty{-4}{\kb}\). Simulation parameters: $T = \qty{1100}{\kelvin}$, $a\mathrm{O_2} = 0.2$, \([\ch{Acc'}] = \qty{8.43e24}{\per\cubic\meter}\), and  $\Delta H_\mathrm{a}=-1$~eV.}
  \label{fig:concentration_along_SCL_and_GB_v3_and_v4}
\end{figure*}

\subsection{Point-defect concentrations in space-charge layers}\label{sec:associates_in_grain_boundary_simulation}

In order to treat the grain boundary and the space-charge layers, we assume the standard two-phase model,\cite{Bingham.1989, Jamnik.1995,McIntyre.2000,Dabringhaus.2003,Tschoepe.2004,DeSouza.2009} in which a grain-boundary phase of finite width ($w^\mathrm{c}$) is sandwiched between two semi-infinite bulk phases. The thermodynamic and kinetic properties of point defects, such as standard defect chemical po-tentials and defect diffusivities, are thus assumed to display one value in the bulk phase and in the space-charge layers all the way up to the core and a different value in the grain‐boundary core. This emphasises the fact that space-charge regions refer to bulk lattice with (strongly) modified defect concentrations. We assume also that the grain boundary is flat and laterally homogeneous, so that a one-dimensional treatment of the electric potential (in the direction $y$ normal to the grain boundary) is sufficient. Furthermore, we assume that space-charge formation is driven by the preferential formation of oxygen vacancies in the grain-boundary core,\cite{McIntyre.2000, Tschoepe.2004, DeSouza.2009} that is, the standard chemical potential of oxygen vacancies is lower in the grain-boundary core than in the bulk, \hbox{$\Delta\mu^\standardstate_\mathrm{v} = \mu^{\standardstate, \mathrm{c}}_\mathrm{v} - \mu^{\standardstate, \mathrm{b}}_\mathrm{v} < \num{0}$;}  $\Delta\mu^\standardstate_\mathrm{def}$  is set to zero for all other defects for simplicity (this constraint can be relaxed;\cite{DeSouza.2009,Usler.2024c} for this study, $\Delta\mu^\standardstate_\mathrm{v} < 0$ is the minimum requirement to guarantee a positive space-charge potential and thus cation-vacancy accumulation in the space-charge zones). Furthermore, we assume that electrochemical potentials $\tilde{\mu}_\mathrm{def}$ can be defined for point defects in the bulk and in the grain-boundary core, and that they take a (modified) Fermi--Dirac form (see below). And we assume that in electrochemical equilibrium, the electrochemical potentials of mobile defects are constant throughout the system.

In this study, a single, temperature-independent value of the thermodynamic driving energy for space-charge formation was used, $\Delta\mu^\standardstate_\mathrm{v} = -2$~eV. This value, it is noted, is consistent with the results of atomistic simulations of selected grain-boundaries in \ch{SrTiO3},\cite{Astala.2002, Uberuaga.2012, Ramadan.2016, Behtash.2018, Mutter.2023} and with values of $\Delta\mu^\standardstate_\mathrm{v}$ required in continuum models to produce physically reasonable space-charge potentials and to explain experimental data quantitatively.\cite{DeSouza.2009} Furthermore, with $\Delta\mu^\standardstate_\mathrm{v}(=\mu^{\standardstate, \mathrm{c}}_\mathrm{v} - \mu^{\standardstate, \mathrm{b}}_\mathrm{v})$ being the difference between two individual point-defect formation free energies, it has been suggested\cite{DeSouza.2019} that a limiting value is probably half an order of magnitude, $\Delta\mu^\standardstate_\mathrm{v}/\unit{\electronvolt} =-10^{0.5} \approx -3.2$, and the taken value is well within this range.

It follows that the Poisson equation to be solved [cf. eqn \eqref{eq:electroneutrality_schottky}] is
\begin{equation}
  -\varepsilon_\mathrm{0}\varepsilon_\mathrm{r} \dfrac{\mathrm{d}^{2} \phi}{\mathrm{d} y^{2}} = e \left( - [\ch{Acc'}] + 2 [\ch{v_{O}^{**}}] - 2 [\ch{v_{Sr}^{''}}] - [\ch{e^{'}}] + [\ch{h^{*}}] \right), \label{eq:Poissongleichung}
\end{equation}
where $\phi$ is the electric potential and $\varepsilon_\mathrm{0} \varepsilon_\mathrm{r}$ is the dielectric permittivity of \ch{SrTiO3}.\cite{Maier.1985} The boundary conditions are $\nabla \phi^\mathrm{b} = 0$ and $\phi^\mathrm{b} = 0$. All five (charged) defects are assumed to be sufficiently mobile to achieve electrochemical equilibrium. To obtain the relationship between local concentration and electrostatic potential, we assume, as noted above, electrochemical equilibrium and a modified Fermi--Dirac form of the electrochemical potential of a defect. The modification is required to take account of various species sharing the same sublattice. For the \ch{Sr} sublattice, for example, \ch{v_{Sr}^{''}} and \ch{(v_{Sr}v_{O})^{x}} need to be taken into account. This leads, for instance, to the electrochemical potential of strontium vacancies being 
\begin{equation}
  \tilde{\mu}_{\ch{v_{Sr}^{''}}}(y) = \mu_{\ch{v_{Sr}^{''}}}^{\standardstate} + k_\mathrm{B}T \cdot \ln\left( \dfrac{[\ch{v_{Sr}^{''}}](y)}{N_{\ch{{Sr}}} - [\ch{v_{Sr}^{''}}](y) - [\ch{(v_{Sr}v_{O})^{x}}](y)}\right) + z_{\ch{v_{Sr}^{''}}} e \phi (y) \label{eq:chempot_general_expression}.
\end{equation}
Here, $N_{\ch{Sr}}$ is the site density of \ch{Sr}, and $z_{\ch{v_{Sr}^{''}}} = -2$ is the relative charge of a strontium vacancy. Rearrangement of eqn~\eqref{eq:chempot_general_expression} yields 
\begin{equation}
  [\ch{v_{Sr}^{''}}](y) = \dfrac{N^\mathrm{gb}_\mathrm{Sr} \cdot [\ch{v_{Sr}^{''}}]^\mathrm{b} \cdot \exp\left({ \dfrac{+2e \phi(y) }{k_\mathrm{B}T} }\right) }{N^\mathrm{b}_\mathrm{Sr} + [\ch{v_{Sr}^{''}}]^\mathrm{b} \cdot \exp\left({ \dfrac{+2e \phi(y) }{k_\mathrm{B}T} }\right) - [\ch{v_{Sr}^{''}}]^\mathrm{b}} \label{eq:c_vsr_full_expression}\\
\end{equation}

\begin{figure*}[h!]
  \centering
  \includegraphics[width=\textwidth]{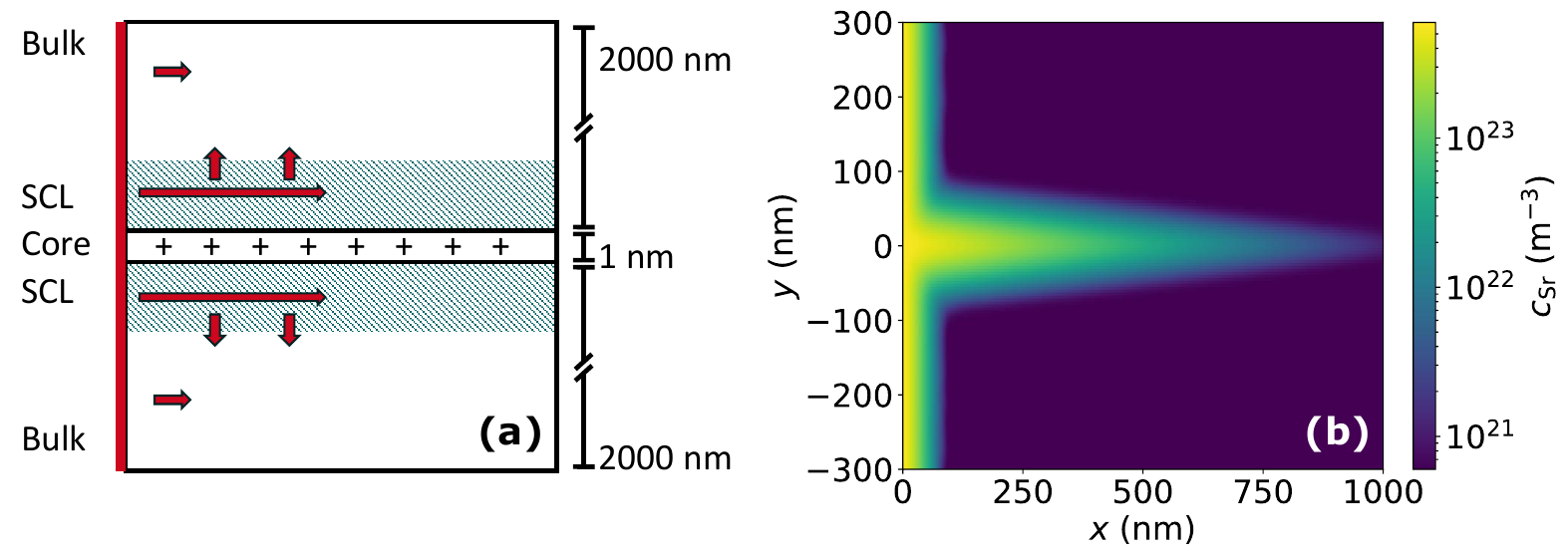}
  \caption{(a) Two-dimensional simulation cell of two grains separated by a grain-boundary core of finite width. Sr diffusion takes place, as indicated by the red arrows, from a constant source (on the left-hand side) into the bulk, and along the space-charge layers and into the bulk (diffusion in the core is negligible). (b) 2D plot of the Sr concentration in the cell after a diffusion anneal, showing enhanced diffusion along the grain-boundary region.}
  \label{fig:2D_simulation_cell_schematic_plus_data}
\end{figure*}

\noindent The equivalent expression for the concentration of oxygen vacancies, including the thermodynamic driving energy for space-charge formation $\Delta \mu^\standardstate_\mathrm{v}$ reads
\begin{align}
  [\ch{v_{O}^{**}}](y) = \dfrac{ [\ch{v_{O}^{**}}]^\mathrm{b} \cdot \exp\left({\dfrac{-2 e \phi(y) - \Delta \mu^\standardstate_\mathrm{v}}{k_\mathrm{B}T}} \right) \cdot (N_\mathrm{O} - [\ch{(v_{Sr}v_{O})^{x}}](y) )  }{ N_\mathrm{O}^\mathrm{b} - [\ch{(v_{Sr}v_{O})^{x}}]^\mathrm{b} + [\ch{v_{O}^{**}}]^\mathrm{b} \cdot \exp\left( {\dfrac{-2 e \phi(y) - \Delta \mu^\standardstate_\mathrm{v}}{k_\mathrm{B}T}} \right) - [\ch{v_{O}^{**}}]^\mathrm{b}} \label{eq:c_vo_full_expression} 
\end{align}
Expressions for electron and hole concentrations follow the standard Maxwell–Boltzmann form $c_\mathrm{def} = c^\mathrm{b}_\mathrm{def} \exp\left(z_\mathrm{def} e \phi(y) / k_\mathrm{B}T\right)$.\cite{Usler.2024}

The spatial variation in the associate concentration does not enter the Poisson equation, since these species are neutral. An expression is required, however, to calculate local strontium diffusivity according to eqn \eqref{eq:diffusivity_sum}. It reads
\begin{equation}
  [\ch{(v_{Sr}v_{O})^{x}}](y) = \dfrac{[\ch{v_{Sr}^{''}}](y)}{\dfrac{[\ch{v_{Sr}^{''}}]^{\mathrm{b}}}{[\ch{(v_{Sr}v_{O})^{x}}]^{\mathrm{b}}} \cdot \exp\left(\dfrac{2 e (\phi(y))}{k_\mathrm{B}T} \right)} \label{eq:c_ass_concentration} 
\end{equation}

\begin{table}[b!]
\small
  \caption{\ Parameters used in this study to model the interfacial defect chemistry of \ch{SrTiO3}.}
  \label{tbl:gb_model_parameters}
  \begin{tabular*}{0.48\textwidth}{@{\extracolsep{\fill}}lll}
    \hline
    Parameter & Value & Units \\
    \hline
    $\Delta \mu^\standardstate_\mathrm{v}$ & \num{-2} & \unit{\electronvolt} \\
    $N\mathrm{_O^{gb}}$ & \num{2.5e27} & \unit{\per\cubic\meter} \\
    $N\mathrm{_{Sr}^{gb}}$ & \num{2.5e21} & \unit{\per\cubic\meter} \\
    $w^\mathrm{c}$ & \num{1e-9} & \unit{\meter} \\
    \hline
  \end{tabular*}
\end{table}

The calculated electric potential and defect concentrations across a grain boundary at $T = \qty{1100}{\kelvin}$ are plotted in Fig.~\ref{fig:concentration_along_SCL_and_GB_v3_and_v4} for different $\Delta S_\mathrm{a}$. In the space-charge layers, oxygen vacancies and electron holes are depleted; and acceptor-dopant cations, isolated strontium vacancies and electrons are accumulated. The concentration of associates shows no significant change within the space-charge layers: such defects are charge neutral; also, the depletion of oxygen vacancies and the accumulation of strontium vacancies fixes through $K_\mathrm{a}$ [eqn~\eqref{eq:associate_formation_equilibrium}] the associate concentration. Since the concentration of acceptor dopants is significantly higher than that of isolated strontium vacancies in the space-charge layer, it is their behaviour that governs the extent of the space-charge layer and the degree to which charged mobile defects are accumulated or depleted. Since [\ch{Acc^'}] is the same for all three cases shown in Fig.~\ref{fig:concentration_along_SCL_and_GB_v3_and_v4}, the electric potential profile across the boundary is the same, too.

\section{Obtaining the Grain-Boundary Diffusion Product}\label{sec:Simulating_grain_boundary_diffusion}

To investigate the diffusion of Sr tracer cations along the space-charge layers at a grain boundary, we consider a two-dimensional cell consisting of two $\ch{SrTiO3}$ grains, sandwiching a grain-boundary core of finite width. Cations may diffuse into the cell from a constant source on one side normal to the grain boundary, as shown in Fig.~\ref{fig:2D_simulation_cell_schematic_plus_data}(a).  

The simulation is executed in three steps. First, bulk concentrations of defects in thermodynamic equilibrium are calculated for given $T$, $a\mathrm{O}_2$, [\ch{Acc'}], $\Delta H_\mathrm{a}$ and $\Delta S_\mathrm{a}$ (Sec.~\ref{sec:Defect_model_SrTiO3_bulk_defects}). Then, the local electrostatic potential (and local defect concentrations) are obtained by solving Poisson's equation for given $\Delta\mu^\standardstate_\mathrm{v}$, $N\mathrm{_O^{gb}}$, $N\mathrm{_{Sr}^{gb}}$ and $T$  (Sec.~\ref{sec:associates_in_grain_boundary_simulation}). From the position-dependent concentrations of \ch{v_{Sr}^{''}} and \ch{(v_{Sr}}\ch{v_{O})^{x}}, the local Sr diffusivity $D_\mathrm{Sr}(y)$ within the simulation cell (i.e.~bulk and space-charge layers) is calculated from eqn~\eqref{eq:diffusivity_sum}; the two defect diffusivities are kept at their bulk values, since the space-charge layers refer to bulk lattice (with altered defect concentrations). We set the cation diffusivity within the core to a negligible level, $D_\mathrm{Sr}^\mathrm{c} = 10^{-6} \cdot D_\mathrm{Sr}^\mathrm{b}$, so that all observed effects can be attributed to the space-charge zones rather than the core. (Also, if transport in the grain-boundary core were to be included correctly, one would need the individual activation barriers for defect migration along the specific diffusion direction in that specific grain-boundary core.) Third, the diffusion equation, 
\begin{equation}
  \dfrac{\partial [\ch{{Sr}}] (x,y) }{\partial t} = \nabla \cdot \left\{ D_\mathrm{Sr} (y) \nabla [\ch{{Sr}}] (x,y) \right\}, \label{eq:diffusion_equation}
\end{equation}
is solved for the strontium tracer cations for diffusion in a semi-infinite medium with zero initial concentration from a constant source. For each temperature, a diffusion time $t$ was calculated, so that a constant characteristic bulk diffusion length of $L_{D} =\sqrt{D^\mathrm{b}_\mathrm{Sr} \, t} = \qty{20}{\nano\meter}$ was obtained. Steps two and three were carried out by means of finite-element-method (FEM) calculations (COMSOL\textregistered Multiphysics\textregistered, Ver. 6.1, Stockholm, Sweden). The result of one such three-step simulation is the two-dimensional distribution of the \ch{Sr} tracer cations shown in Fig.~\ref{fig:2D_simulation_cell_schematic_plus_data}(b).

\begin{figure}[h!]
  \centering
  \includegraphics[width=\columnwidth]{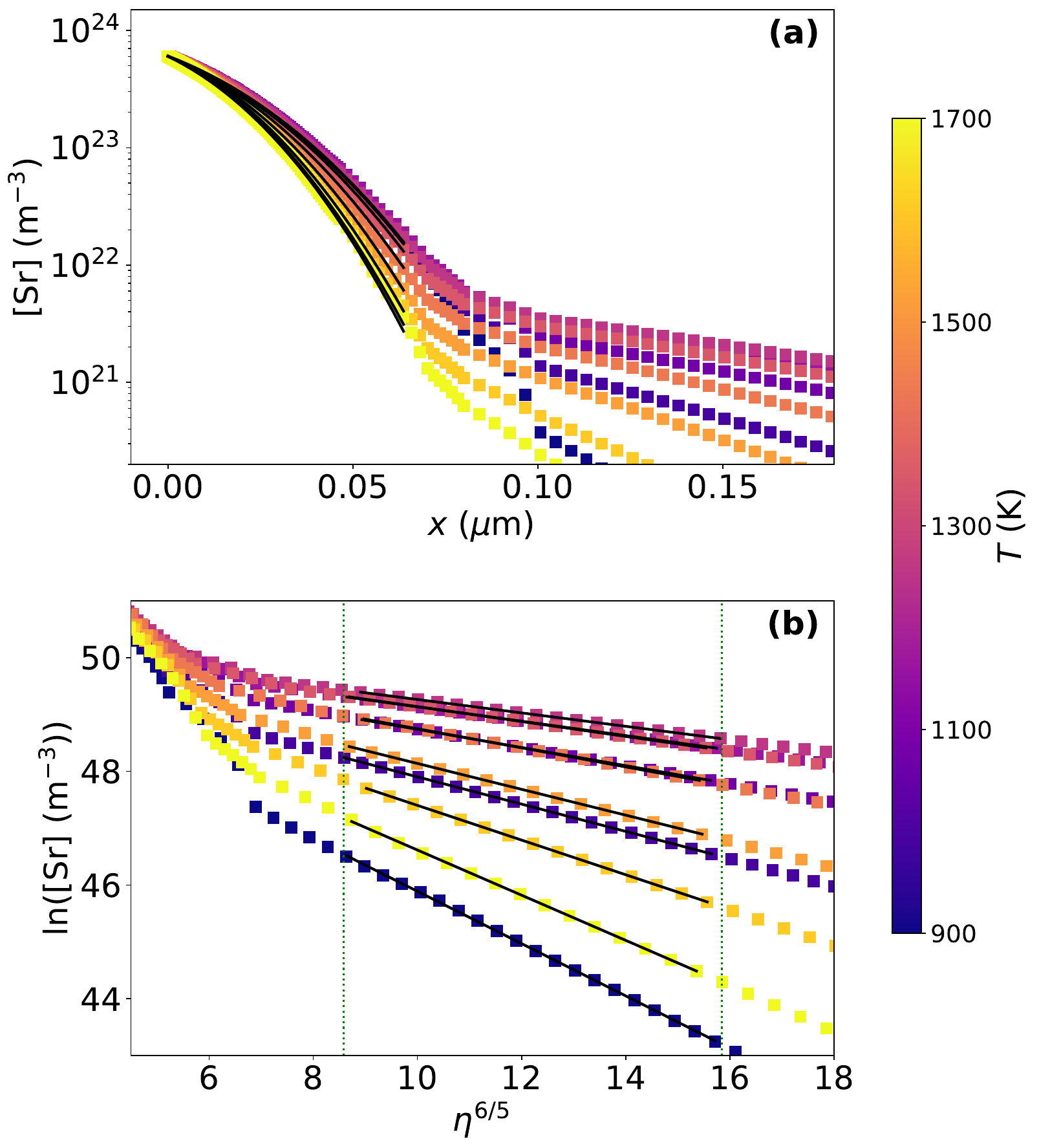}
  \caption{Strontium tracer diffusion profiles obtained for different temperatures $900 \leq T/\mathrm{K} \leq 1700$ for the bicrystal geometry; diffusion times were chosen so that the bulk diffusion length $L_D$ was the same for all temperatures. (a) Strontium tracer concentration versus depth, together with error function fits to the first (bulk) feature. (b) Concentration as a function of \(\eta^{6/5}\), with linear fit to the second feature, the grain-boundary `tail'. Simulation parameters: $\Delta S_\mathrm{a} = \qty{-2}{\kb}$, \([\ch{Acc'}] = \qty{8.43e24}{\per\cubic\meter}\), \(\Delta H_\mathrm{a} = \qty{-1}{\electronvolt}\), \(\Delta \mu_\mathrm{v}^{\standardstate} = \qty{-2}{\electronvolt}\), \(N\mathrm{^{gb}_{O}} = \qty{2.5e27}{\per\cubic\meter}\), \(N\mathrm{^{gb}_{Sr}} = N\mathrm{^{gb}_{O}} \cdot 10^{-6}\).}
  \label{fig:depth_profile_with_fit}
\end{figure}

The calculated two-dimensional concentration distribution was projected onto the $x$ axis to obtain a one-dimensional depth profile of the strontium tracer concentration, similar to an experimental depth profile obtained by Secondary Ion Mass Spectrometry (SIMS). The simulation parameters were chosen to produce diffusion profiles referring to Harrison B-type diffusion kinetics\cite{Harrison.1961} with two distinct features evident (see Fig.~\ref{fig:depth_profile_with_fit}), corresponding to slow diffusion in the bulk and faster diffusion along the grain boundary and slow bulk diffusion out of it. 

For bulk diffusion, the appropriate solution to the diffusion equation, given the chosen initial and boundary conditions, is a complementary error function,\cite{Crank.1975} and unsurprisingly this mathematical form describes the first feature well, as shown in Fig. \ref{fig:depth_profile_with_fit} (a), yielding $D^\mathrm{b}_\mathrm{Sr}$. Values obtained in this manner are very close to the $D^\mathrm{b}_\mathrm{Sr}$ values used as input [taken from Fig.~\ref{fig:full_model_D_and_H} (a)], with deviations being less than \qty{3}{\percent}.

The second feature exhibits linear behaviour in a plot of $\ln{[\ch{Sr}]}$ versus $\eta^{6/5}$ [Fig.~\ref{fig:depth_profile_with_fit} (b)], where $\eta = y / L_D$, as expected for faster grain-boundary diffusion.\cite{Levine.1960, LeClaire.1963} Its analysis follows the procedure proposed by Chung and Wuensch,\cite{Chung.1996} in which the slope of the second feature for $6 \leq \eta \leq 10$ is used to obtain the grain-boundary diffusion product $a^\mathrm{gb}D^\mathrm{gb}_\mathrm{Sr}$ according to 
\begin{equation}
    a^\mathrm{gb} D^\mathrm{gb}_\mathrm{Sr} = (D^\mathrm{b}_\mathrm{Sr})^{3/2}\, t^{1/2} \left[ 10^{A} \left( - \dfrac{\partial \ln [\ch{Sr}]}{\partial \eta^{6/5}} \right)^{B} \right]. \label{eq:Chung_Wuensch_equation_agb_Dgb}
\end{equation}
where $a^\mathrm{gb}$ is the grain-boundary half-width. Values of $A$ and $B$, which depend on the value of the slope, were derived by Chung and Wuensch.\cite{Chung.1996} In order to obtain diffusion profiles with a sufficiently pronounced grain-boundary feature, we only considered data for which $\beta \geq 10$, where
\begin{equation}
 \beta = \frac{D^\mathrm{gb}}{D^\mathrm{b}} \frac{a^\mathrm{gb}} {\sqrt{D^\mathrm{b}  t}}, 
 \label{eqn:beta}
\end{equation}
provides a measure of the magnitude of the grain-boundary diffusivity relative to that of the bulk. 

\begin{figure}[b!]
  \centering
  \includegraphics[width=1.0\columnwidth]{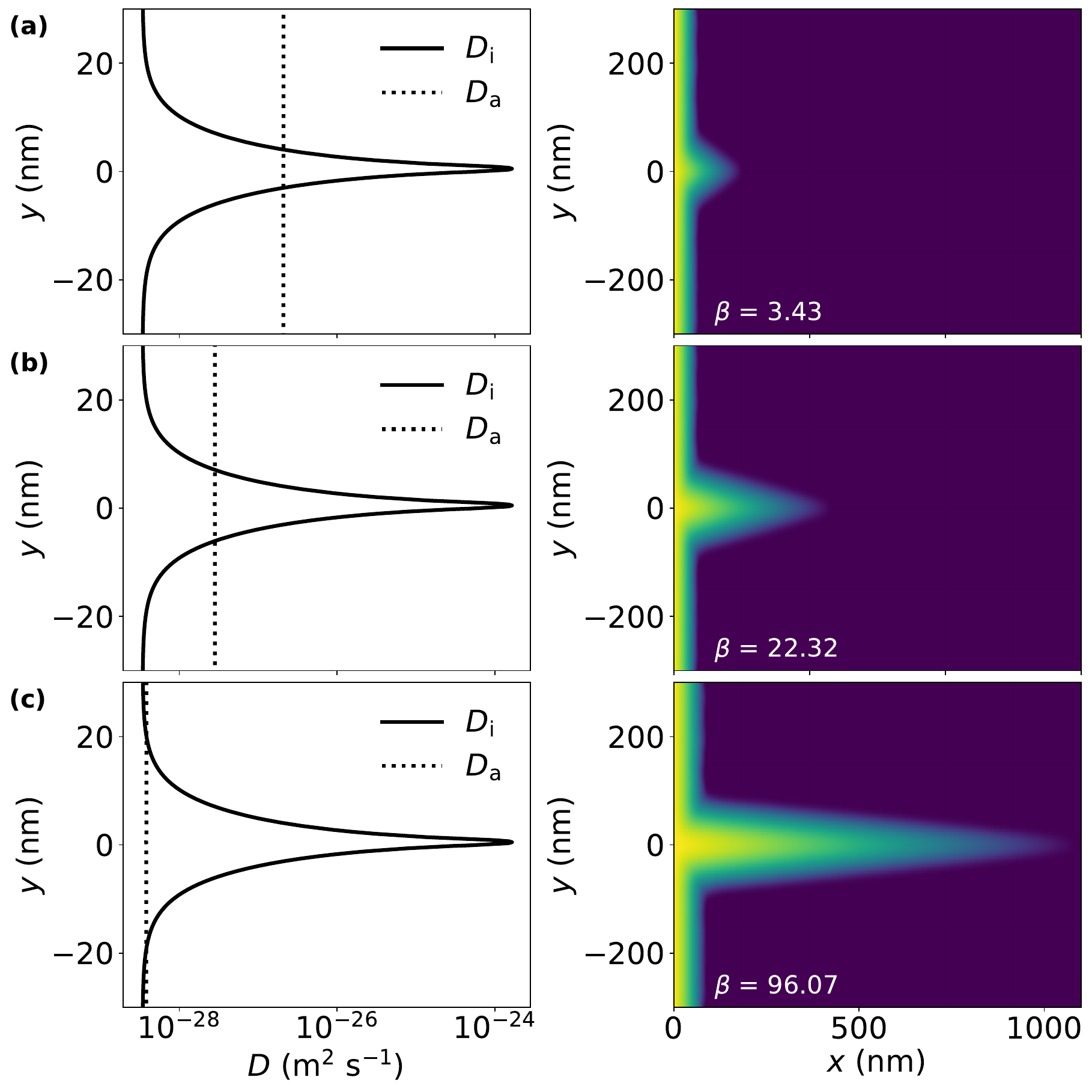}
  \caption{Local Sr diffusion coefficients across the grain boundary for isolated vacancies $D_\mathrm{Sr,i}(y)$ and defect associates $D_\mathrm{Sr,a}(y)$ (left), and 2D concentration plots for different diffusion times $t$ that gave constant bulk diffusion length $L_D$ (right). (a)~\(\Delta S_{\text{a}} = \qty{0}{\kb}\), (b) \(\Delta S_{\text{a}} = \qty{-2}{\kb}\), (c) \(\Delta S_{\text{a}} = \qty{-4}{\kb}\). $T = \qty{1200}{\kelvin}$.}
  \label{fig:Dy2D}
\end{figure}

In Fig.~\ref{fig:Dy2D}, we plot local cation diffusivities across the grain boundary for three different $\Delta S_\mathrm{a}$ values and the corresponding 2D concentration plots that result from diffusion anneals (of differing times, to give the same $L_D$). One sees that the greater the difference between $D^\mathrm{b}_\mathrm{Sr,a}$ and the maximum diffusivity in the space-charge layer, $D\mathrm{_{Sr,i}^{max}}$ [this difference increasing in the order (a), (b), (c)], the stronger the preferential penetration of the diffusant is, as indicated by the higher $\beta$. The degree to which \ch{v_{Sr}^{''}} are accumulated remains, however, the same; it is the bulk diffusivity that changes, owing to changes in [\ch{(v_{Sr}v_{O})^{x}}] produced by differing $\Delta S_\mathrm{a}$. Given that $\beta\propto t^{-1/2}$ [see eqn~\eqref{eqn:beta}], one can recognise that, with increasing diffusion time, preferential penetration along the space-charge layers at grain boundaries will become less evident. Thus, it is conceivable, for sufficiently long diffusion times, that no preferential diffusion is detected because of $\beta \ll 10$, even though $D\mathrm{_{Sr,i}^{max}} > D^\mathrm{b}_\mathrm{Sr,a}$. That is, the lack of a second feature in a diffusion profile cannot be taken as evidence of no space-charge layers being present at grain boundaries.

\section{Results and Discussion}

\begin{figure}[th!]
  \centering
  \includegraphics[width=\columnwidth]{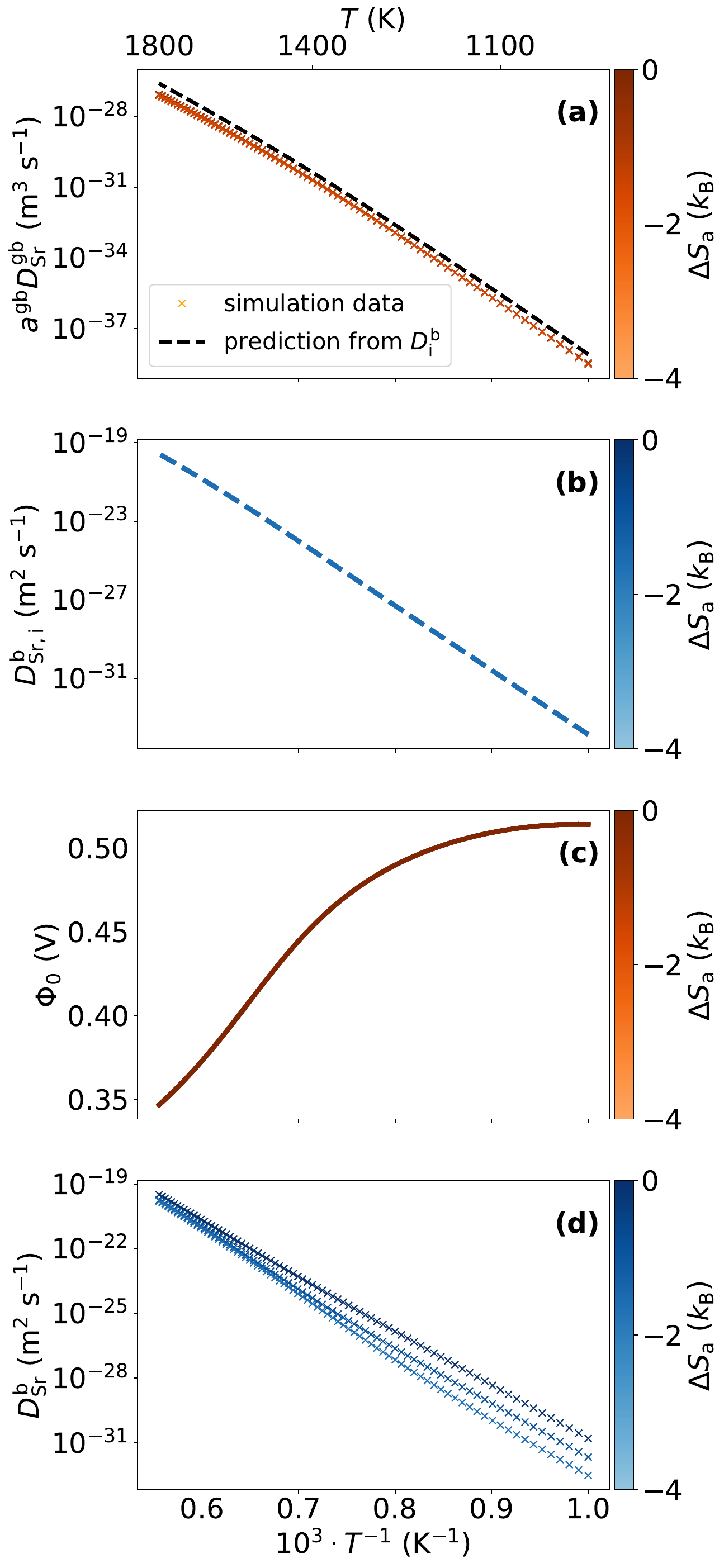}
  \caption{Selected bulk and grain-boundary properties obtained from the simulations, plotted as a function of inverse temperature. (a) Grain-boundary diffusion product from the simulations, $a^\mathrm{gb}D^\mathrm{gb}$, and predicted with eqn~\eqref{eq:Parras_aD_pred_model_adapted}, $a^\mathrm{gb}D^\mathrm{gb}_\mathrm{pred, i}$. (b) Bulk diffusion coefficient due solely to the diffusion of isolated strontium vacancies,  $D^\mathrm{b}_\mathrm{Sr, i}$; (c) space-charge potential $\Phi_{0}$; (d) bulk diffusion coefficient $D^\mathrm{b}_\mathrm{Sr}$ from error function fit. NB: for all but (d), data for different $\Delta S_\mathrm{a}$ are plotted but lie on top of each other.}
  \label{fig:SCP_D_b_gb_superplot}
\end{figure}

The grain-boundary diffusion products obtained from the simulations for a range of temperatures are plotted in Fig.~\ref{fig:SCP_D_b_gb_superplot}~(a) for selected values of $\Delta S_\mathrm{a}$. The data all fall together on a single line; data for lower $\Delta S_\mathrm{a}$ (not shown)  also fall on this line. For the same selected $\Delta S_\mathrm{a}$, the bulk diffusivities of \ch{Sr} are significantly different, especially at lower temperatures [Fig.~\ref{fig:SCP_D_b_gb_superplot}~(d) reproduced from Fig.~\ref{fig:full_model_D_and_H}(a)]. The reason for this difference is that $D^\mathrm{b}_{\ch{Sr}}$ has contributions from associated and isolated strontium vacancies, whereas $a^\mathrm{gb}D^\mathrm{gb}$ is dominated by the isolated strontium vacancies, for it is these defects that are strongly accumulated in the space-charge layers.

Since space-charge layers are regions of modified bulk defect chemistry, one can understand the observed behaviour in terms of (i) the relevant bulk defect concentrations and (ii) the degree of their modification in the space-charge layers. Thus, we first show in Fig.~\ref{fig:SCP_D_b_gb_superplot}~(b) the contribution of the \ch{Sr} diffusivity arising solely from isolated strontium vacancies, that is $D_\mathrm{i} [\ch{v_{Sr}^{''}}] / [\ch{Sr_{Sr}^{x}}]$, see eqn~\eqref{eq:diffusivity_sum}; and since, as noted previously, $[\ch{v_{Sr}^{''}}]^\mathrm{b}$ is unaffected by the association equilibrium, we find that $D_\mathrm{Sr,i}^\mathrm{b}$ is also independent of $\Delta S_\mathrm{a}$. In Fig.~\ref{fig:SCP_D_b_gb_superplot}~(c) we show the space-charge potential $\Phi_{0}$, and in this case, as also noted previously, since the accumulation of the acceptor-dopant cations, $[\ch{Acc^{'}}](y)$, governs the screening within the space-charge layers, the space-charge potential is also independent of $\Delta S_\mathrm{a}$. In this way, we understand why all the $a^\mathrm{gb}D^\mathrm{gb}$ fall together and are independent of $\Delta S_\mathrm{a}$. We also understand that the important bulk defect concentration in this case is $[\ch{v_{Sr}^{''}}]$, since it is these defects whose concentration is enhanced in the space-charge layers, giving rise to faster diffusion along the grain boundary.

This understanding can also be placed on a quantitative level. Parras and De Souza\cite{Parras.2020} derived an approximate expression for the grain-boundary diffusion product (in the case of diffusion along space-charge layers) in terms of the bulk diffusion coefficient, the Debye length $\ell_\mathrm{D}$ and the space-charge potential, 
\begin{equation}
  a^\mathrm{gb}D^\mathrm{gb}_\mathrm{pred, i}(T) = 2  \ell_\mathrm{D}(T) f_{\ell} \cdot D^\mathrm{b}_\mathrm{Sr, i}(T) \cdot \exp\left( - \dfrac{f_\mathrm{P} z_\mathrm{v_\mathrm{Sr}} \mathrm{e} \cdot \Phi_\mathrm{0} (T)}{k_\mathrm{B} T} \right),  \label{eq:Parras_aD_pred_model_adapted}
\end{equation}
with the empirical parameters $f_{\ell} = \num{0.205}$ and $f_\mathrm{P} = \num{0.854}$. This equation essentially expresses the fact that faster grain-boundary diffusion along space-charge layers can be considered to be accelerated (i) over a distance $a^\mathrm{gb}$, which is some fraction $f_{\ell}$ of twice the Debye length, and (ii) by an amount given by the exponential term, which refers to an effective average (hence: $f_\mathrm{P}$) over the entire space-charge layer. With the data in Fig.~\ref{fig:SCP_D_b_gb_superplot}~(b) and (c) we thus calculated $a^\mathrm{gb}D^\mathrm{gb}_\mathrm{pred, i}$, and we compare the values in Fig.~\ref{fig:SCP_D_b_gb_superplot}(a). We find good agreement between the values from the simulations and the values predicted with eqn~\eqref{eq:Parras_aD_pred_model_adapted}, the difference being less than a factor of $3$. In this way the main points from above are confirmed quantitatively.

In principle, the Debye length, \hbox{\(\ell_\mathrm{D} = \sqrt{\varepsilon_\mathrm{0} \varepsilon_\mathrm{r} k_\mathrm{B} T / (2 e^{2} [\ch{Acc'}]^\mathrm{b})}\)}, is expected to display a weak dependence on temperature in the present case, not only explicitly through $T$ but also through the Curie--Weiss behaviour\cite{Maier.1985} of $\varepsilon_\mathrm{r}$ of \ch{SrTiO3}. In practice, for the temperature range $1000 \leq T/K \leq 1800$, $\ell_\mathrm{D}$ is calculated to decrease from \qty{5.13}{\nano\meter} to \qty{5.09}{\nano\meter}. The grain-boundary half-width, that is, the effective distance over which diffusion is accelerated on one side of the boundary, according to eqn~\eqref{eq:Parras_aD_pred_model_adapted}, is thus $a^\mathrm{gb} = 2 \ell_\mathrm{D}f_{\ell}\approx 2.1$~nm. This yields, for example at $T=\qty{1300}{\kelvin}$ with $\Delta S_\mathrm{a}=\qty{-2}{\kb}$, an effective average of $D^\mathrm{gb}_\mathrm{Sr}=\qty{3.7e-24}{\meter\squared\per\second}$ (with $D^\mathrm{b}_\mathrm{Sr}=\qty{1.6e-26}{\meter\squared\per\second}$).

\begin{figure}[t!]
  \centering
  \includegraphics[width=\columnwidth]{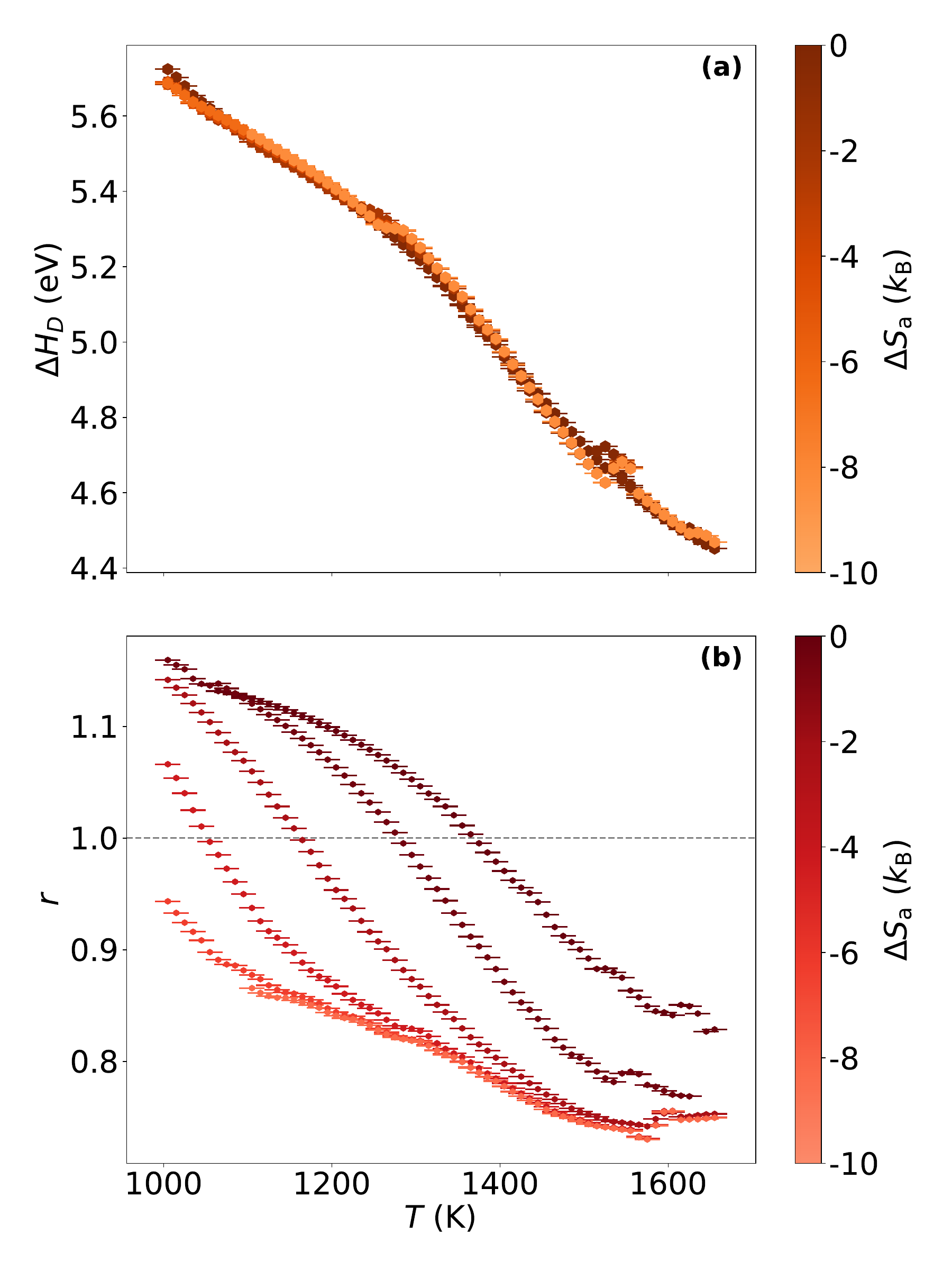}
  \caption{(a) Activation enthalpy of strontium grain-boundary diffusion, $\Delta H\mathrm{^{gb}_{Sr}}$, determined as a function of temperature over a rolling interval of \qty{40}{\kelvin} for different entropies of associate formation. (b) Ratio $r$ of the activation enthalpies of diffusion in the grain boundary and in the bulk as a function of temperature for different entropies of associate formation. $r=1$ is indicated by a dashed grey line.}
\label{fig:bulk_and_gb_Enthalpies_comparison_v3_and_v4}
\end{figure}

Now we come to the main question of this study, and the possibility of $r > 1$. In Fig.~\ref{fig:bulk_and_gb_Enthalpies_comparison_v3_and_v4} (a) we plot the activation enthalpy of the grain boundary diffusion product obtained for a rolling interval of 40 K. $\Delta H^\mathrm{b}_{D} (T)$ decreases monotonically with increasing temperature, and is not a function of $\Delta S_\mathrm{a}$, reflecting the behaviour of Fig.~\ref{fig:SCP_D_b_gb_superplot}~(a). In contrast, $\Delta H^\mathrm{b}_{D}$ [Fig. \ref{fig:full_model_D_and_H}~(b)] tends to go through a maximum with increasing temperature and is a function of $\Delta S_\mathrm{a}$. Combining the data of Fig.~\ref{fig:bulk_and_gb_Enthalpies_comparison_v3_and_v4}~(a) and Fig.~\ref{fig:full_model_D_and_H}~(b) we obtain $r$ as a function of temperature for various values of $\Delta S_\mathrm{a}$ [Fig. \ref{fig:bulk_and_gb_Enthalpies_comparison_v3_and_v4}~(b)]. The datasets for the most negative values of $\Delta S_\mathrm{a}$ correspond essentially to the case of isolated strontium vacancies dominating diffusion in the bulk, and these data show $r < 1$. The datasets for the higher values of $\Delta S_\mathrm{a}$ contain a predominant contribution to $D^\mathrm{b}_{\ch{Sr}}$ from the \ch{(v_{Sr}v_{O})^{x}} associates, especially at lower temperatures, and it is under these conditions that we find $r>1$. For $\Delta S_\mathrm{a} / k_\mathrm{B} = -2$, for instance, $r$ values up to $\num{1.15}$ are attained for the temperature range considered. Beyond the fact that $r > 1$ is possible, we emphasise the fact that $r$ varies with temperature.

The behaviour of $r$ in Fig.~\ref{fig:bulk_and_gb_Enthalpies_comparison_v3_and_v4}~(b) suggests that even higher values of $r$ may be attained at even lower temperatures, and thus it raises the question of whether there is a limit to $r$ for this model with the defect-chemical parameters of Tables~\ref{tbl:defect_model_parameters} and \ref{tbl:gb_model_parameters}. For the case of a single diffusion mechanism,\cite{Parras.2020} $r \leq 1$ was established as the physical limit based on the behaviour of $\Phi_{0}(T)$. In the present case, we can extend the arguments of Parras and De Souza\cite{Parras.2020} to the more complicated model. The first part is that the highest possible value of $\Delta H^\mathrm{gb}_{D}$ is $\Delta H^\mathrm{b}_{D,\mathrm{i}} \approx \qty{6.5}{\electronvolt}$ (i.e.~that of the isolated strontium vacancies), and it corresponds to $r = 1$ for the isolated vacancies. The second part is that $r > 1$ is generated by $\Delta H^\mathrm{b}_{D}$ being lower than $\Delta H^\mathrm{b}_{D,\mathrm{i}}$ through a dominant contribution from the \ch{(v_{Sr}v_{O})^{x}} associates. And the lowest value for 
$\Delta H^\mathrm{b}_{D}$ is $ \qty{4.9}{\electronvolt} $. Thus, putting the two parts together we obtain --- for this system with these parameters --- an upper limit of $r \leq 1.3$.

To test this model, experimental studies are necessary for comparison. And in performing such studies, the range of temperatures considered is extremely important: as we have found, $r$ is a function of temperature and tends to exceed unity at lower rather than higher temperatures. Common temperature ranges for experimental cation diffusion studies in titanate perovskites are in the range $\num{1200} \leq T/K \leq \num{1500}$.\cite{Meyer.2003,Bak.2004,Gomann.2005,Gries.2020,Koerfer.2008} In this temperature range, our simulation for $\Delta S_\mathrm{a}/\unit{\kb} = \num{-2}$ predicts $r>1$. But in our simulations we have a high density of data points, and this will generally not be the case in experiment. Indeed, if one only considers five data points, $T/\mathrm{K} = \{1200, 1250, 1300, 1350, 1400\}$, one would obtain $\Delta H_{D}^\mathrm{gb} = \qty{5.22\pm0.04}{\electronvolt}$ and $\Delta H_{D}^\mathrm{b} = \qty{5.31\pm0.05}{\electronvolt}$, hence a ratio of $r = \num{0.98\pm0.01}$ (for $\Delta S_\mathrm{a} / \unit{\kb} = \num{-2}$). Thus, an even lower temperature range is necessary for an experimental study to be able to demonstrate $r > 1$. For five temperatures in the range $\num{1100} \leq T/\mathrm{K} \leq \num{1300}$, we predict $\Delta H_{D}^\mathrm{gb} = \qty{5.42\pm0.03}{\electronvolt}$ and $\Delta H_{D}^\mathrm{b} = \qty{5.08\pm0.03}{\electronvolt}$, and hence $r = \num{1.07\pm0.01}$.

\section{Concluding Remarks}

Using an established set of defect chemical parameters for the bulk of acceptor-doped \ch{SrTiO3}; using physically reasonable values of $\Delta S_\mathrm{a}$ and $\Delta H_\mathrm{a}$ for the formation of \ch{(v_{Sr}v_{O})^{x}} associates; using defect diffusivities that show higher values for the associate than for the isolated vacancy; using a standard thermodynamic model of space-charge formation; and taking a reasonable value for the thermodynamic driving energy of space-charge formation $\Delta \mu^{\standardstate}_{\mathrm{v}}$; we find for the faster diffusion of cations along the space-charge layers at a grain boundary than in the bulk ($D^\mathrm{gb} > D^\mathrm{b}$) that $\Delta H_{D}^\mathrm{gb} > \Delta H_{D}^\mathrm{b} (=r > 1)$ is indeed possible under certain conditions.

With respect to the previous study of Parras and De Souza,\cite{Parras.2020} in which only isolated cation vacancies were considered, and in which $r \leq 1$ was found to be the physical limit, the main extension in this study is the consideration of two types of cation vacancies, isolated and associated, with different defect diffusivities. And as we demonstrated, it is this extension that gives rise to the possibility of $r>1$. Our treatment employed on purpose a relatively simple defect chemical model, ignoring, for example, defect ionisation equilibria, and a relatively simple model of the interface, ignoring, for example, local strain fields, as this allowed us to show clearly why $r>1$ arises.

Finally, we comment on the applicability of this model to other systems. We presume that the general behaviour is applicable to the diffusion of $A$ cations in all $AB\mathrm{O}_3$ perovskites, first because the formation of $\mathrm{(v_\mathit{A}v_{O})}$ associates is highly likely given the opposite relative charges of the two constituent defects, and second, because the higher diffusivity of the associate is expected as a characteristic of the $AB\mathrm{O}_3$ perovskite structure. Indeed, for \ch{BaTiO3}, \ch{(v_{Ba}v_{O})} associates are predicted to be bound,\cite{Erhart.2007} and they have recently been predicted\cite{Koerfer.2026} to have higher diffusivities than isolated \ch{v_{Ba}^{''}}. That said, $\mathrm{(v_\mathit{A}v_{O})}$ associates are not the only associates containing cation vacancies that may form in a perovskite, and the presence of significant concentrations of $\mathrm{(v_\mathit{A}v_\mathit{B})}$ associates\cite{Bonkowski.2024,Koerfer.2026} or $\mathrm{(v_\mathit{A}v_\mathit{B}v_{O})}$ clusters\cite{Schulz.2003,Belova.2007} may give rise to even richer cation diffusion behaviour.

\section*{Acknowledgements}

This study has received funding from the Deutsche Forschungsgemeinschaft (DFG, German Research Foundation) within project 463184206 (SFB 1548, FLAIR: Fermi Level Engineering Applied to Oxide Electroceramics). Discussions with Matthew J. Wolf are acknowledged.





\bibliography{rsc} 
\bibliographystyle{rsc} 

\end{document}